\documentclass{edi}

\usepackage{makeidx}    
\usepackage{graphicx}   
\usepackage{listings}   
\usepackage{setspace}   
\usepackage{hyperref}   

\makeindex              

\begin{document}
\doublespacing
\chapter{}
\date{}
\title{Beyond Classroom: \\ Making a Difference in \\ Diversity in Tech}
\author{Barbora Buhnova}
%
%
\maketitle

With all the opportunities and risks that technology holds in connection to our safe and sustainable future, it is becoming increasingly important to involve a larger portion of our society in becoming active co-creators of our digitalized future---moving from the passenger seat to the driver seat. Yet, despite extensive efforts around the world, little progress has been made in growing the representation of certain communities and groups in software engineering. This chapter shares one successful project, called Czechitas, triggering a major social change in Czechia, involving 1\,000+ volunteers to support 50\,000+ women on their way towards software engineering education and career.

\vspace{-3mm}
\section{Introduction}
\vspace{-2mm}

The past decade has witnessed the emergence of hundreds of initiatives around the world supporting various underrepresented groups on their pathway towards software engineering, whether connected to universities~\cite{MinervaBestPractices}, companies~\cite{CompaniesWomenInTech}, or run as independent non-profit organizations~\cite{OrganizationsWomenInTech}. Although the initiatives often start with a great vision and high volunteering commitment, after a few years into the activities, it becomes challenging to sustain the volunteering energy and commitment in face of the very slow progress towards the better. In those moments, the success cases by others can be what helps us keep going.

\begin{figure}[b!]
\centering
\includegraphics[width=\textwidth]{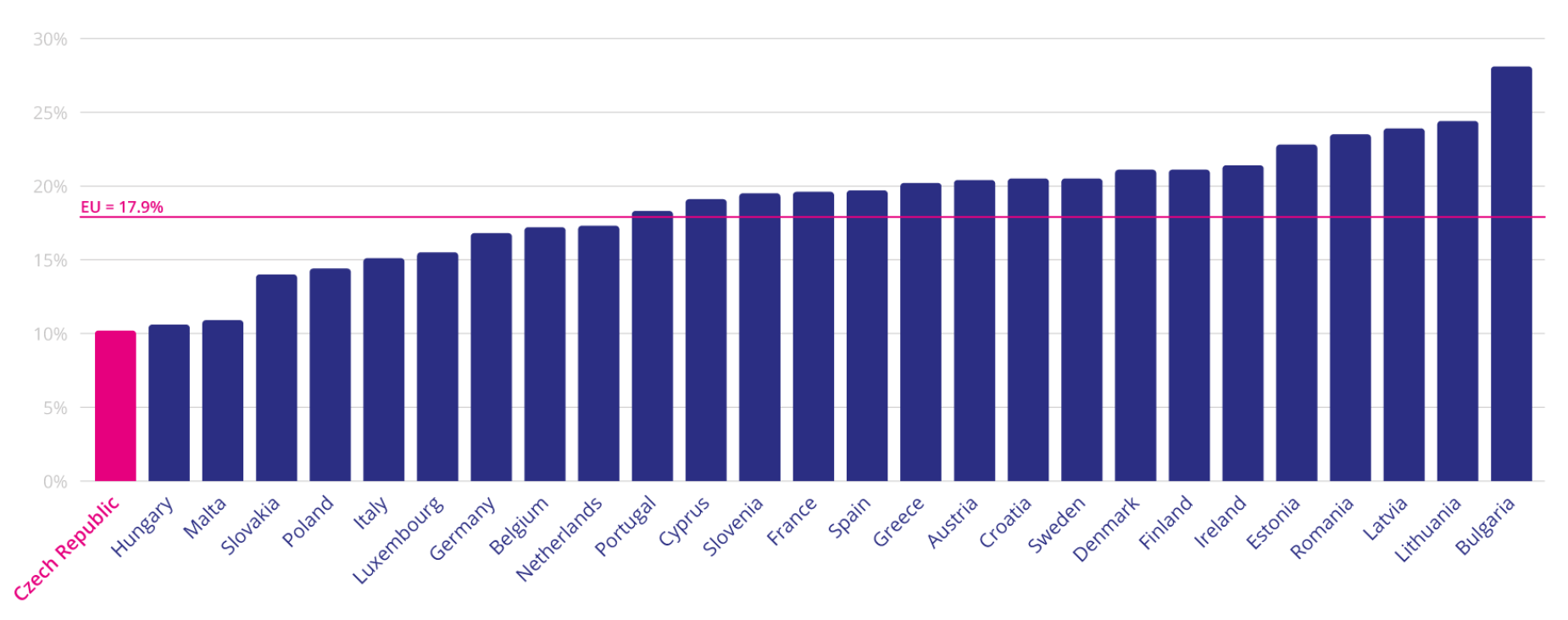} \vspace{-5mm}
\caption{Women ICT Professional (Eurostat, 2019 data)~\cite{EurostatICTSpecialists}.}
\label{fig:Women-ICT-professional-2019}
\end{figure}

The initiative featured in this chapter, called Czechitas~\cite{CzechitasAnnualReport2021}, started in 2014 in Czechia, with a simple idea to bring tech closer to girls and girls closer to tech, in reaction to the strong under-representation of women in tech in the country (see Figure \ref{fig:Women-ICT-professional-2019}). The prompt snowball effect helped us to build a community around the joint vision to empower and encourage girls and women to engage in computing education and career transition, and to show them that software engineering is an interesting career direction that is not necessarily difficult nor limited to one gender. Initially established to provide women in Czechia with an opportunity to put their hands on programming, it now contributes to a major social change in the country.

Over time, Czechitas has become a movement that has attracted a strong community of tech-professional volunteers (over 1\,000) and companies (over 100), and given rise to a portfolio of women-tailored courses in various areas of software engineering, such as programming, web development, mobile app development, data science, cybersecurity or testing (over 1\,300 courses delivered so far). We have influenced over 50\,000 women (over 30\,000 via live events and over 20\,000 via online tutorials) who graduated from our courses to use their new tech skills to change their education path or advance their careers.

\begin{tcolorbox} \tcbfontsize{1.2}
\textbf{Czechitas Mission:} We inspire, train and guide new talents towards stronger diversity and competitiveness in tech.
\end{tcolorbox}

Thanks to the success of our education activities with hundreds of events a year (each receiving more registrations than its capacity), we have become recognized as the leading platform in Czechia actively addressing gender diversity in tech. In this chapter, we share the lessons we learned about the low representation of women in tech, effective strategies in supporting women on their way to software engineering, discuss the ingredients that helped us succeed, the obstacles and challenges we faced, and the progress yet to be made.




\vspace{-3mm}
\section{Why are There so Few Women in Tech?}
\vspace{-2mm}

Across Europe, only 19.1\% of tech professionals are women (according to 2021 data)~\cite{EurostatICTSpecialists}, with Czechia being the last on the list. The major reasons behind the trend in our region according to our recent study (with 70\% of participants from Czechia and Germany)~\cite{happe2021frustrations} are:

\begin{enumerate}

\item \textbf{Access.} The first hole in the leaky pipeline on girls' pathway towards software engineering is linked to the \emph{missing access to encouragement and support}, together with the \emph{missing access to suitable education} that would be able to build on the interests of girls that often span across multiple disciplines.

\item \textbf{Stereotypes.} The ability to \emph{see herself as a software engineer} is then challenged by the \emph{perception of the software engineering} as a field not leading to a purpose the girl would like to dedicate her future to. Often, the close family and friends step-in in this moment to direct girls away from software engineering \emph{with the intention to protect them} from a future where they cannot really imagine the girls becoming successful. Interestingly, the intentions are meant well, to protect the girls, which shows how crucial it is to help parents (and mainly mothers) to understand that software engineering can be a great career choice for their daughters.

\item \textbf{Confidence.} The next hole on the leaky pipeline comes when girls find themselves in the classroom, often \emph{surrounded by more-experienced learners} (typically boys). For the little girls who often excel in other subjects, it can be hard to fall in the category of a slow novice learner. The girls often mention frustrations of \emph{low self-efficacy}, \emph{inadequacy} and \emph{missing experience of success} in presence of a classroom dynamic being monopolized by the earlier technology adopters.

\item \textbf{Sense of Belonging.} The girls who resist through the earlier three challenges and find themselves on the education pathway towards software engineering, find themselves in classrooms surrounded predominantly by boys. While this is a comfortable environment for some, many in the study reported \emph{not feeling comfortable to express themselves}, facing \emph{sexism or unwanted attention} and \emph{missing relatable role models and mentors}, which led them to reconsider whether this is the environment they would be willing to spend the rest of their lives in.

\item \textbf{Feeling Valued.} The last hole in the leaky pipeline challenges the women who entered software engineering careers, as some of them emphasize the struggle of not feeling valued at workplace. The reasons are different for the women with \emph{stereotypical talent spectrum} (that matches the talent spectrum typical among their men colleagues, typically being very technical) and \emph{non-stereotypical talent spectrum} (bringing not-that-common talents to the table, typically more multidisciplinary and human-oriented). While the first group feels \emph{"tired of proving them wrong"}, the second group feel frustrated from \emph{their strengths viewed as second class} and from \emph{missing appreciation}.

\end{enumerate}


\vspace{-3mm}
\section{Supporting Women on their Way to Tech}
\vspace{-2mm}

\begin{figure}[b!]
\centering
\includegraphics[width=0.5\textwidth]{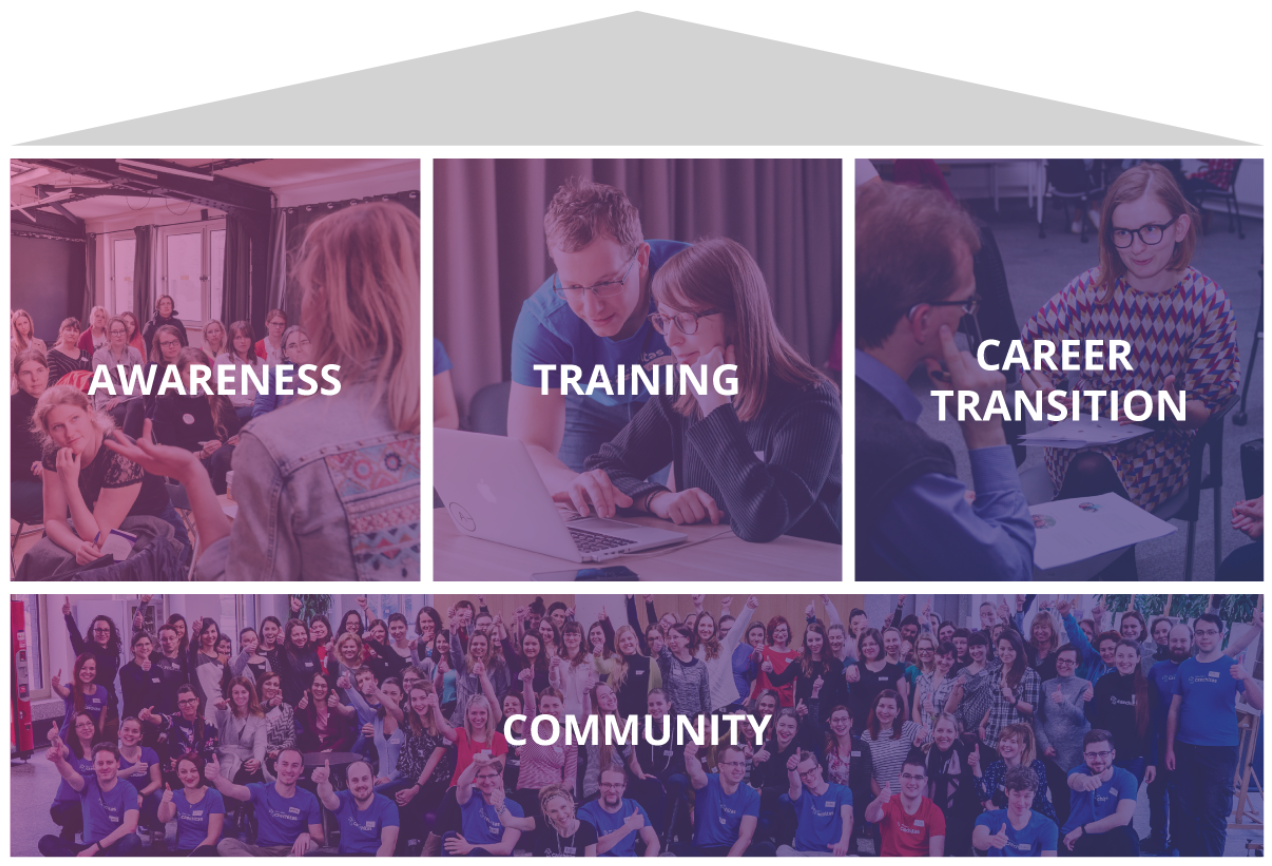}
\caption{The Pillars of Czechitas Activities.}
\label{fig:CzechitasPillars}
\end{figure}

In Czechitas, we understand that plumbing the leaky pipeline can hardly be done by isolated and uncoordinated efforts. This section discusses the interlinked pillars of our activities (see Figure \ref{fig:CzechitasPillars}), listing examples of the activities and events we delivered in 2022.

\vspace{-3mm}
\subsection{Czechitas Pillar I -- Awareness}
\vspace{-2mm}

One of the crucial success factors for a change towards improving gender balance in software engineering is the actual understanding that we are in a disbalanced state that further reinforces itself due to the factors discussed earlier. The efforts towards encouraging women to join software engineering cannot make a difference unless the society, education system and corporate environment welcomes and supports the change (understanding it as a push towards the real equilibrium, not a push out of it). 

In Czechitas, we are investing substantial effort in awareness around the topic. In 2022 alone, we participated in over 20 conferences and panel discussions, gave numerous interviews in TV, radio and other media, organized talks to students and teachers at high schools, and to tech professionals in our partner companies. We were visible with a booth at 15 festivals and family days across Czechia. Over 2022, Czechitas was mentioned in 508 articles, reaching major part of Czech population. In 2021, we also launched a Czechitas podcast, which in 2022 reached over 14\,676 listens. Furthermore, our website was in 2022 visited by 123\,785 unique visitors, and our newsletter was followed by 25\,983 subscribers.

The next step in raising awareness among the general public is to make it as easy as possible to get the first exposure to coding in a fun, enjoyable and community way. To this end, we for instance organize an Advent Christmas Coding campaign (following the tradition of an advent calendar, in which instead of a sweet treat, each day holds a coding assignment along a story of bringing Mr. Gingerbread home for Christmas), which is being followed by hundreds of people. Furthermore, in collaboration with the Ministry of Education, Youth and Sports, we e.g. co-organized the \#DIGIEDUHACK hackathon. And in collaboration with Czech universities, we run the Czechitas Thesis Award to give visibility to exceptional bachelor theses authored by girls. All these activities typically repeat every year.

\begin{figure}[b!]
\centering
\includegraphics[width=\textwidth]{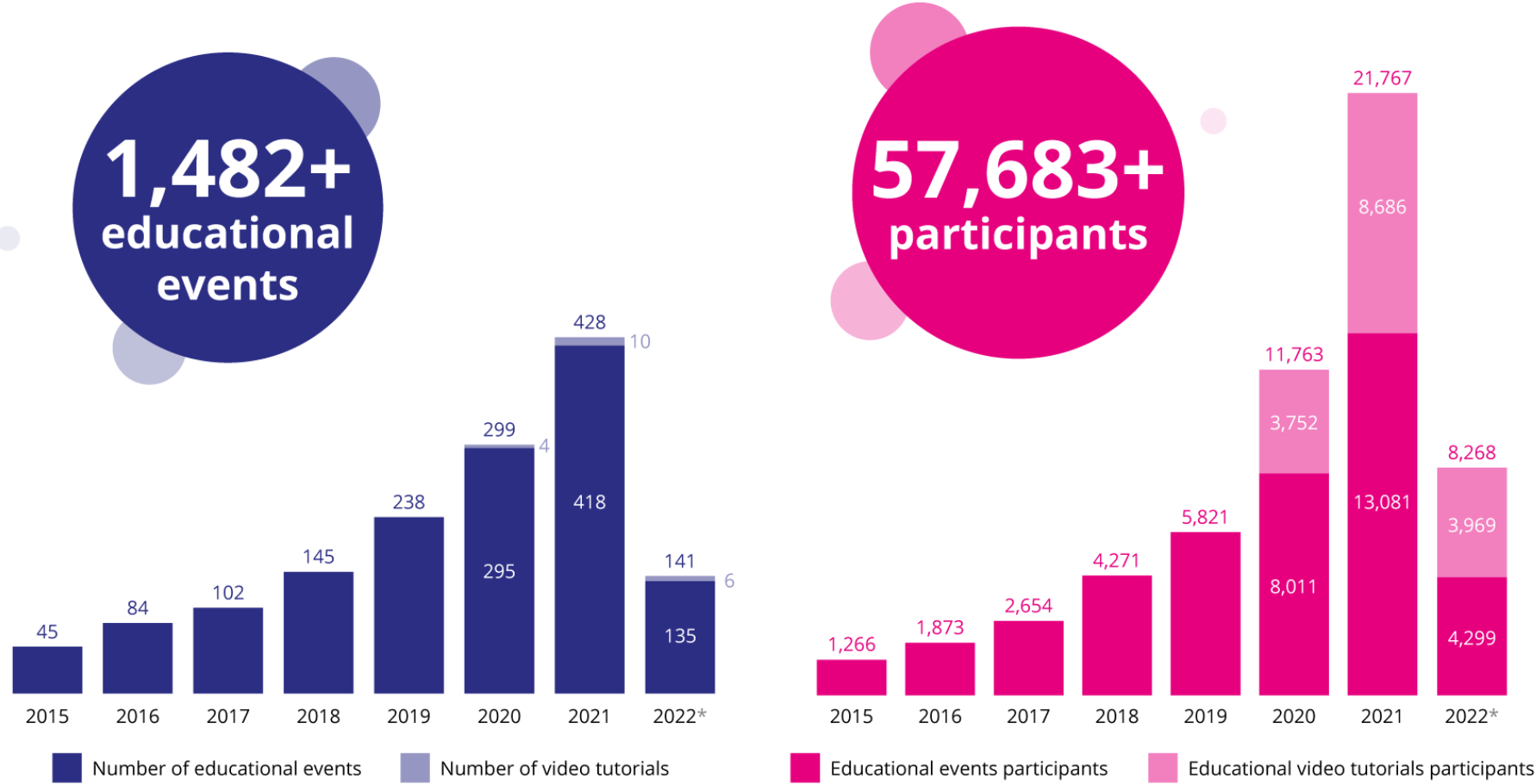}
\caption{Czechitas-participation (Data as of June 2022).}
\label{fig:Czechitas-participation}
\end{figure}

\vspace{-3mm}
\subsection{Czechitas Pillar II -- Training}
\vspace{-2mm}

Since the start of our activities in 2014, we are improving the education design of our courses to reflect the needs of our audience---women and girls who are very often later technology adopters or career changers---with an emphasis on providing suitable first contact with software engineering, creating safe and supportive environment for novice learners, accommodating differences in the learning speed of each student, building self-confidence, and supporting sustaining long-term interest, which we also publish~\cite{buhnova2020girl,happe2021effective}. In 2022, we delivered 242 live software-engineering courses with 15\,316 participants, with the courses around web development and data science scoring as the most popular ones.

Although most of the training is targeted to women and girls, we are also investing in training elementary-school and high-school teachers (irrespective of gender). And some mixed-gender activities were organized also for children (7 week-long summer camps in the summer of 2022, besides others) and high-school kids, although in case of high schools, it is already important to offer also girl-only courses (3 week-long summer schools for high-school girls were given in 2022). Besides, training courses for mixed audience are also provided on events such as Family Days (we were present at over 20 such events in 2022). 

\vspace{-3mm}
\subsection{Czechitas Pillar III -- Career Transition} 
\vspace{-2mm}

As many women in our community intend to enter software engineering as their future profession, some of our activities are intentionally designed to facilitate this journey, whether software engineering is to become their first job or they intend to change their career~\cite{buhnova2019assisting}.

In cooperation with our partner companies, we have identified three career pathways that appear to be the most suitable entry points to software engineering in Czechia. These are \emph{(1) web development} (including courses on JavaScript, React, HTML/CSS, Bootstrap, Git, UX design, and others), \emph{(2) data analytics} (including courses on Python, databases, SQL, statistics, Power BI, and others), \emph{(3) testing} (including courses on requirements engineering, agile processes, manual testing, issue tracking, regression testing, smoke testing, basics of automated testing, browsers, API, databases, version control, and others).

For the three directions, we have developed a complex career-transition support within so-called Digital Academies. A Digital Academy is a four-month program for a group of 30 women (and involving around 5-15 partner companies), which besides individual courses covering the topics outlined above and taking place 3-4 times a week (evenings on working days, full days on weekends) includes also pairing of the students with mentors from the companies to support them in developing their own projects, a hackathon, career support, and further events offered by the partner companies. In 2022, we have run 10 Digital Academies across four major cities in Czechia, with over 60\% of the graduates receiving a job offer within three months from graduating from the academy.

To facilitate the career transition also for the women who opt to customize their training journey (not attending a Digital Academy), our career consultants provide hundreds of career consultations each year (327 in 2022), and we twice a year organize a Czechitas Job Fair, where our graduates can meet the representatives of our partner companies (each Job Fair attended by about 350 graduates and 30 companies).

\vspace{-5mm}
\subsection{Czechitas Foundation -- Community}
\vspace{-2mm}

The foundation that supports all our activities is the community, which involves the participants and graduates of our courses, tech professionals who teach with us, mentors, course facilitators, and our partner companies. The fact that many members in our community are men helps us not only engage more tech-professional allies in our vision, but also influence a more supportive environment in tech companies where our graduates land. To support the blending of the community and increasing the sense of belonging of our graduates also in the mixed-gender environment, we regularly engage in organization of Tech Meet-ups and Hackathons, as well as informal CzechiPubs that regularly take place in 10 different cities across Czechia.


\vspace{-5mm}
\section{Making a Difference}
\vspace{-2mm}

\begin{figure}[h]
\centering
\includegraphics[width=\textwidth]{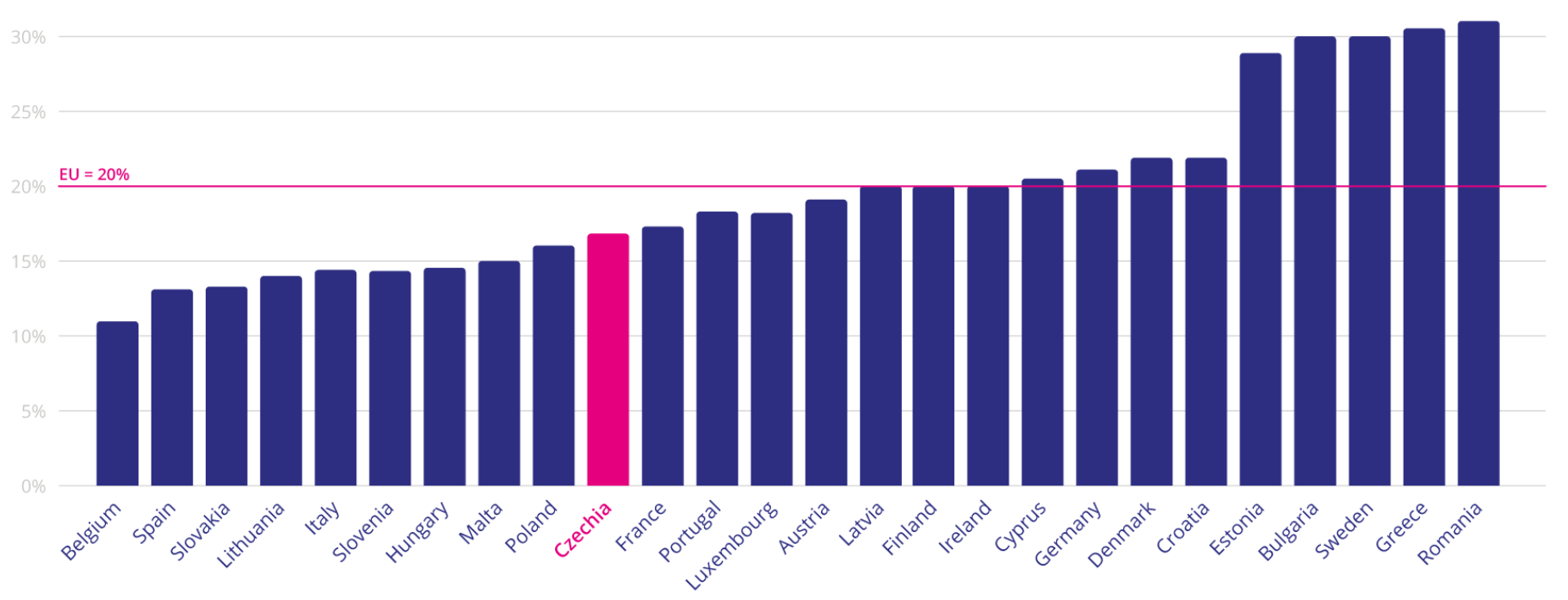}
\caption{Women ICT Students (Czech Statistical Office, 2021 data)~\cite{CSU-ICTStudents2021}.}
\label{fig:Women-ICT-students-2021}
\end{figure}

The positive influence of Czechitas activities in Czechia is already visible in the shifted perception of software engineering as an education pathway and career choice to be considered by any gender. That not only motivates many girls to consider software engineering in their choice of a university study field (with the representation of women among ICT students changing from 12\% in 2016 to 17\% in 2021 in Czechia~\cite{EurostatICTStudents2016,CSU-ICTStudents2021}, moving the country closer to the European average, see Figure \ref{fig:Women-ICT-students-2021}) but is likely also having secondary influence on all who so far hesitated to join software engineering.  

\vspace{-5mm}
\subsection{What Helped us Succeed}
\vspace{-2mm}

Building Czechitas was only possible thanks to a coordinated effort of hundreds of people (90 employees and over 1,000 volunteers). Over the past eight years of our existence, we came to understand the ingredients without which this would not be possible: 

\begin{itemize}

\item \textbf{Great leadership and love for what we do} is giving us the sense of purpose, energy and direction, holding us together and keeping us focused. Mentors from partner companies and beyond have been of great help to guide us through the design of our leadership and expansion strategy.

\item \textbf{Visual and playful communication} is giving us the fresh flavour of fun and joy that we all (students as well as trainers and volunteers) enjoy joining even after a tiring day at school or work. The informal and visually attractive communication helps us to share the love for our brand.

\item \textbf{Community and sense of belonging} is crucial for connecting those who strive to learn with those who strive to share and teach, and those who want to support the connection. It helps our student to feel home and make it easier for them to keep going even when learning gets hard.

\item \textbf{Inclusive environment and encouragement} makes it safe for our students to make mistakes and experience success, have the opportunity to exchange knowledge, collaborate, and get personalized feedback and guidance. Specific strategies and interventions we have developed to support novice learners and their self-efficacy have been key in this direction~\cite{buhnova2020girl}.

\item \textbf{Knowledge and understanding} is crucial for us to design our activities with insight into the frustrations steering women away from software engineering~\cite{happe2021frustrations} and effective strategies to support girls and women in tech education~\cite{happe2021effective} and career transition~\cite{buhnova2019assisting}. We invest our time in sharing the lessons we have learned~\cite{buhnova2020girl, happe2021frustrations, buhnova2019assisting}, and learning from other initiatives from across the world (e.g., within the EUGAIN network, see https://eugain.eu/). 

\item \textbf{Creating and sharing stories} helps us to inspire our students, bring them closer to relatable role models, and to give them hope and confidence that with some work and dedication, a transition into software engineering is possible. The stories (each featuring an inspiring woman who changed her career towards tech) are published in our blog, communicated via social networks, and used in media articles. These women inspire others as speakers and panelists in our events, and as guests in Czechitas Podcast. 

\item \textbf{Sustainable financial model} helps us to sustain a team employed to run the organization. The model stands on financial participation of the students, partner companies, foundations and individual donors, with an intention to reach out also to the government level in the future. The most crucial pillar of our financial sustainability is the partner companies, which are beside their yearly partnership contributions (depending on the level of partnership) helping us to cover certain costs (e.g., offering their office spaces for events, motivating their employees to volunteer as mentors), and opening doors towards further funding opportunities (e.g. with global foundations connected to their company).

\end{itemize}


\vspace{-3mm}
\subsection{Obstacles and Challenges we Faced}
\vspace{-2mm}

As any organization that has substantially outgrown its own plans and expectations, Czechitas has undergone numerous changes and readjustments over its course of existence. And although we are trying to publish the effective setup that works for us now~\cite{buhnova2020girl, buhnova2019assisting, buhnova2019women}, our first steps were highly organic and experimental, which was key to learning what works for the context we were in. With our enthusiasm and "always yes" spirit, we walked many paths that we failed and rolled back, but we also faced numerous obstacles and challenges that we withstood. 


\begin{itemize}

\item \textbf{Scaling the organization.}
Turning a non-profit start-up into a scale-up is a challenge on its own, as the means for achieving stability are different from traditional companies -- besides the discussed financial stability, also in terms of sustained volunteering involvement and brand building. We needed to learn to manage the mix of the innovative and largely self-sacrificing founding community with the necessary systematic and organized spirit of new employees. We needed to learn to prioritize and say no to some activities that the team felt strongly for. 

\item \textbf{Being misunderstood.}
As a large organization, we needed to learn to communicate our mission well so that it is not misunderstood, knowing that anything that damages the brand may sink the whole boat. Namely, we needed to help our partner companies understand what level of expertise is realistic to achieve in our students, help our students understand what time investment and commitment it takes to change direction towards tech, and help our society understand why our focus on women is key to the success of our society as a whole.



\item \textbf{Quantifying the impact of our activities.}
One of the important challenges that we are still facing is our ability to quantify the impact of our individual interventions and activities, as it is difficult to isolate the effects of each one of them. More so that the impact is often very subtle and propagates over long periods of time (e.g., a woman making a few steps towards tech education inspiring her friend to make a major shift towards tech, who then inspires her daughter to study CS at university). So although we have a Data \& Impact team at Czechitas, with substantial data available, the numbers we have (e.g., the number of women who change their career to tech each year) are still only the tip of the iceberg of the real impact we strive for, which is the shift in the collective mindset of the entire society, leading to a sustained change. 

\end{itemize}


\vspace{-3mm}
\subsection{Progress yet to be Made}
\vspace{-2mm}

With the increasing number of Czechitas graduates who are joining software engineering industry, often as very junior (in terms of their software-engineering expertise) and diverse (in terms of their talents and competencies) members, we find it crucial to assist the companies to improve the inclusiveness of their environment to integrate and leverage the new diverse talent.
In 2020, we made the first step towards that goal via designing a \emph{Diversity Awareness Training}, which was since then delivered to over 300 managers (mostly from Central and Eastern Europe) across some of our partner companies. The concepts that have shown to be the most crucial to discuss and understand during these trainings are outlined below: 

\begin{figure}[t!]
\centering
\includegraphics[width=0.8\textwidth]{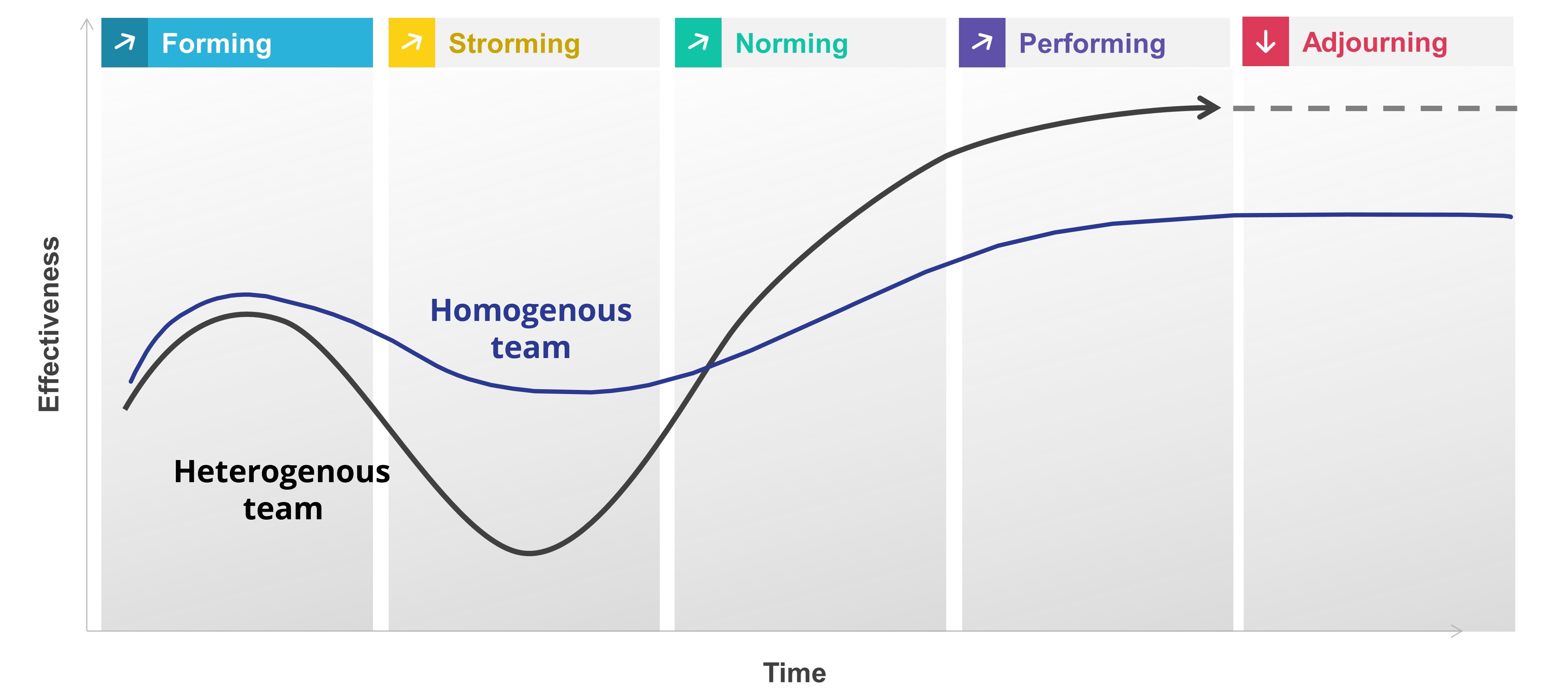}
\caption{Tuckman's Model of Team Dynamics with an illustration of different dynamics observed in homogeneous and heterogeneous teams.}
\label{fig:Tuckman}
\end{figure}

\begin{itemize}

    \item \textbf{Diversity does not come easy, but it pays off.} Avoiding diversity is natural to human individuals, but dangerous to humankind\footnote{Our quote inspired by the statement "Diversity is the new Darwinism" by the Great British Diversity Experiment~\cite{BritishDiversityExperiment2016}.}. The same is true for corporate environment. We need to acknowledge that diverse teams might have a harder time at start (as illustrated with the Tuckman's Model of Team Dynamics in Figure \ref{fig:Tuckman}), but in long-term, diversity is firmly correlated with higher performance~\cite{McKinsey2015,DiversityPerformance2018}.
    
    \item \textbf{We too often lose talented people by missing the talent in them.} We are all talented, in many diverse ways. It is the task of the manager to recognize and direct the talent towards team success. The fact that a person uses a different talent spectrum (approaches problems and situations differently) does not make them more/less suitable for software engineering as such. There is no such thing as a second-class citizen when it comes to the talents we need in software engineering.
    
    \item \textbf{Biases evolved to help us navigate complexity, but they are not serving us well when making assumptions about the potential in people.} The dark side of biases is that we tend to judge people's potential based on how their talent spectrum matches the talent of already-successful ones. Without realizing that the successful ones embody the skills and conditions that worked when they joined the field (in the past) while we are now choosing the software engineers for the future.

   
    \item \textbf{Connection is built through communication.} There are many unhealthy communication patterns around diversity, which often go against the purpose of making us all feel the sense of belonging. It is important to create safe space, in which we can learn to communicate our differences but also ask about the differences of others. Mistakes are part of that learning, and forgiveness of the mistakes shall be encouraged if the mistakes were done in the process of learning and not repeated blindly. It is important to create a safe space to acknowledge our biases and stop shaming one another for them. 

    \item \textbf{Avoid the quick fixes, remove the barriers instead.} Encourage curiosity about why certain communities are under-represented in software engineering. What are the barriers they face and what can we do to remove them or make their journey lighter in presence of the barriers (e.g. the care-taking on the side of most women)? Avoiding the conversation and looking away from the differences in our experiences might lead the community to assume that the under-representation is the lower-fit problem, which is dangerous because it leads to push-back on any diversity support one might try to introduce. 

    \item \textbf{Change takes time.} Promoting I\&D is more complex than it might seem at first. It is crucial to know how to start to see the first positive effects soon and be able to use them to get more people on board towards promoting I\&D further. Choose your first steps well and invest in them. The investment will pay off.
    
\end{itemize}


\vspace{-3mm}
\section{Conclusion}
\vspace{-2mm}

Making a difference in improving gender balance in software engineering on the scale of the whole country is not easy, but is possible. And it is very rewarding to be part of such a movement. In 2021, the social impact of Czechitas activities was recognized at the European Union level via winning the EU Social Economy Award (over 180 organizations nominated) in the Digitalisation and Skills category, and in 2022 winning the global Equals in Tech Award (155 organizations nominated) in the Skills category. We hope our example can inspire others, which is also why we are eager to share the lessons learned from our journey.



\bibliographystyle{plain}
\bibliography{refs}


\section{Acknowledgement}

This chapter was made possible thanks to the great dedication and support of the entire Czechitas team. Besides, it has been supported by the COST Action CA19122 -- European Network for Gender Balance in Informatics (EUGAIN).

\end{document}